\documentstyle[12pt]{article}
\def\be{\begin{eqnarray}}
\def\ee{\end{eqnarray}}
\def\ba{\begin{array}}
\def\ea{\end{array}}
\def\p{\phi}
\def\vp{\varphi}
\def\a{\alpha}

\def\eps{\varepsilon}
\def\pa{\partial}
\def\b h{\hbar}
\def\t{\widetilde}
\def\M{{\cal M}}

\def\H{{\cal H}}

\def\K{{\cal K}}
\def\L{{\cal L}}
\def\E{{\cal E}}

\def\T{{\cal T}}
\def\P{{\cal P}}

\def\C{{\cal C}}

\begin{document}
\begin{center}
{\LARGE { Application of the canonical quantization of systems\\
\vskip 0.3cm with curved phase space to the EMDA theory}}
\end{center}

\vskip 1.5cm

\begin{center}
{\bf \large {J.E. Paschalis}$^{\star,}$\footnote{E-mail:
paschali@hermes.ccf.auth.gr}\large {\ \ and \ A.
Herrera--Aguilar}}$^{\star,}$$^{\natural,}$\footnote{E-mail:
aherrera@auth.gr}
\end{center}

\vskip 1cm

\begin{center}
{$^\star$\it
Theoretical Physics Department, Aristotle University of Thessaloniki\\
54124 Thessaloniki, Greece}
\end{center}

\begin{center}
{$^\natural$\it
Instituto de F\'\i sica y Matem\'aticas, UMSNH\\
Edificio C--3, Ciudad Universitaria, Morelia, Mich. CP 58040 M\'exico}
\end{center}

\begin{abstract}
The canonical quantization of dynamical systems with curved phase
space introduced by I.A. Batalin, E.S. Fradkin and T.E. Fradkina
is applied to the four--dimensional Einstein--Maxwell
Dilaton--Axion theory. The spherically symmetric case with radial
fields is considered. The Lagrangian density of the theory in the
Einstein frame is written as an expression with first order in
time derivatives of the fields. The phase space is curved due to
the nontrivial interaction of the dilaton with the axion and the
electromagnetic fields.
\end{abstract}
\noindent PACS numbers:


\newpage
\section{Introduction}
The main idea of I.A. Batalin, E.S. Fradkin and T.E. Fradkina
\cite{bf}--\cite{bff} for the canonical quantization of systems
with curved phase space consists of performing a dimensionality
doubling of the original phase space by introducing a set of new
variables, equal in number to the variables of the original phase
space, so that each one of them is defined as the conjugate
canonical momentum to each original phase variable. Further, the
complete set of variables is subjected to special second class
constraints in such a way that the formal exclusion of the new
canonical momenta reduces the system back to the original phase
space. It turns out that the new phase space is flat and its
quantization proceeds along the lines of \cite{bf2}--\cite{bff2}.

In this paper we shall apply this method to the bosonic sector of
the truncated four--dimensional effective field theory of the
heterotic string at tree level, better known as Einstein--Maxwell
Dilaton--Axion (EMDA) theory. Since this truncated effective
theory contains just massless bosonic modes, we shall consider a
suitable purely bosonic model.

The paper is organized as follows: in Sec. 2 we present a brief
outline of the generalized canonical quantization method for
dynamical systems. We keep the notation and terminology of the
authors. In Sec. 3 we consider the action of the four--dimensional
EMDA theory, perform the ADM decomposition of the metric and write
the Lagrangian density of the matter sector as an expression with
first order in time derivatives of the fields. We further consider
the spherically symmetric anzats and obtain a Lagrangian density
which defines a curved phase space and possesses two irreducible
first class constraints. We continue by canonically quantizing the
resulting EMDA system along the lines of \cite{bf2}--\cite{bff2}
in Sec. 4. In order to achieve this aim, a suitable generalization
of the method has been performed. We sketch our conclusions in
Sec. 5 and, finally, we present some useful mathematical
identities and relationships in the Appendix A.

\section{Outline of the method}

Let there be a dynamical system described by the original
Hamiltonian $H_0=H_0(\Gamma)$ given as a function of $2N$ bosonic
phase variables \be \Gamma^A, \qquad\qquad A=1,2,...,2N,
\label{vars} \ee of certain manifold $\M$.

We assume that the dynamical system is unconstrained. The
extension of the formalism to systems with constraints is
straightforward. The Lagrangian of the system can be presented as
an expression with first order in time derivatives (denoted by a
dot over the variable) of the phase space variables $\Gamma^A$
\cite{jackiw} \be L=a_A(\Gamma)\dot\Gamma^A-H_0(\Gamma).
\label{L}\ee The Euler--Lagrange equations for the Lagrangian
(\ref{L}) are given by the following relations
\be\omega_{AB}(\Gamma)\dot\Gamma^B=\frac{\pa}{\pa
\Gamma^A}H_0(\Gamma), \label{eom}\ee where \be
\omega_{AB}(\Gamma)=\frac{\pa}{\pa
\Gamma^A}a_B(\Gamma)-\frac{\pa}{\pa \Gamma^A}a_A(\Gamma).
\label{wAB}\ee

The nondegenerate tensor $\omega_{AB}(\Gamma)$ defines on the
phase space manifold the covariant components of the symplectic
metric. Also it is antisymmetric and satisfies the Jacobi
identity. The Poisson bracket for any two functions $X(\Gamma)$
and $Y(\Gamma)$ of the phase space variables (\ref{vars}) is
defined as follows:
\be\{X,Y\}=\pa_AX\omega^{AB}\pa_BY,\label{XY}\ee where
$\pa_A\equiv\pa/\pa\Gamma^A$.

Next, the reper field with contravariant components
$h^A\,_a(\Gamma)$ and its inverse $h^a\,_A(\Gamma)$ are introduced
as follows \be
\omega^{AB}(\Gamma)=h^A\,_a\,\omega^{ab}_{(0)}h^B\,_b, \qquad
\omega_{AB}(\Gamma)=h^a\,_A\,\omega_{ab}^{(0)}h^b\,_B, \qquad
a,b=1,2,...,2N;\label{haA}\ee where $\omega^{ab}_{(0)}$ and
$\omega_{ab}^{(0)}$ define a constant symplectic metric which we
will assume that it has the form \be
\omega_{ab}^{(0)}=\left(\ba{cc}0&I_N\cr -I_N&0\ea \right),
\label{sympmetric} \ee where $I_N$ stands for the unit matrix of
dimension $N$.

The covariant derivatives in the phase space are defined as
follows \be\nabla_CV_A\equiv\pa_CV_A-\Delta^D_{CA}V_D, \qquad
\nabla_CV^A\equiv\pa_CV^A+\Delta^A_{CD}V^D,\label{covder}\ee where
$\Delta^D_{CA}\equiv h^D\,_a\,\pa_C\,h^a\,_A$.

The commutator of the covariant derivatives is given by the
following relations \be
[\nabla_A,\nabla_B]=-\Lambda^C_{AB}\nabla_C,\label{Lambdabr}\ee
where \be
\Lambda^C_{AB}\equiv\Delta^C_{AB}-\Delta^C_{BA}.\label{Lambda}\ee

The covariant derivatives of the reper fields and the symplectic
metric are zero. By looking at the description of the classical
dynamics of the system under consideration, it can be verified
that the equations of motion written in terms of the Poisson
brackets (\ref{XY}) can be derived from the action \be
S=\int\left[\Gamma^A\overline\omega_{AB}(\Gamma)\dot\Gamma^B-H_0(\Gamma)\right]dt,
\label{action}\ee where
\be\overline\omega_{AB}(\Gamma)\equiv\int\omega_A(\a\Gamma)\a
d\a.\label{omegabar}\ee

Now we introduce new bosonic variables $\Pi_A$ equal in number
with the initial phase variables $\Gamma^A$. The new variables are
taken as the conjugate canonical momenta to the initial variables.
For any two functions of the $\Gamma^A$ and $\Pi_A$,
$X(\Gamma^A,\Pi_A)$, $Y(\Gamma^A,\Pi_A)$, we define the following
flat Poisson bracket
\be\{X(\Gamma^A,\Pi_A),Y(\Gamma^A,\Pi_A)\}'\equiv\pa_AX\pa^AY-\pa_AY\pa^AX,\label{XY'}\ee
where $\pa^A\equiv\pa/\pa\Pi_A$.

Let us consider the action \be
S'=\int\left[\Pi_A\dot\Gamma^A-H_0(\Gamma)-\Theta_A(\Gamma,\Pi)\lambda^A\right]dt,
\label{action'}\ee where \be
\Theta_A(\Gamma,\Pi)\equiv\Pi_A+\overline\omega_{AB}(\Gamma)\Gamma^B\label{Thetas}\ee
and $\lambda^A$ are Lagrange multipliers.

The flat Poisson bracket for any two of them is given by the
following relation
\be\{\Theta_A,\Theta_B\}'=\omega_{AB}(\Gamma).\label{Thetas'}\ee

The action (\ref{action'}) represents a system with Hamiltonian
$H_0(\Gamma)$ independent of the momenta $\Pi_A$ subjected to
special second class constraints $\Theta_A(\Gamma,\Pi)$. It turns
out that the equations of motion that follow from the action
(\ref{action'}) are equivalent to those derived from the initial
action (\ref{action}) after the formal exclusion of the canonical
momenta $\Pi_A$.

In the case when the dynamical system possesses original
irreducible first class constraints $T'_a(\Gamma)$, $a=1,2,...,m'$
(this is indeed the case for the model we will consider later on),
the expressions (\ref{action}) and (\ref{action'}) are
correspondingly modified as follows \cite{bff}: \be
S=\int\left[\Gamma^A\overline\omega_{AB}(\Gamma)\dot\Gamma^B-H_0(\Gamma)-
T'_a(\Gamma)\lambda'^a\right]dt, \label{actfcc}\ee \be
S'=\int\left[\Pi_A\dot\Gamma^A-H_0(\Gamma)-\Theta_A(\Gamma,\Pi)\lambda^A-
T'_a(\Gamma)\lambda'^a\right]dt. \label{actfcc'}\ee

The next step is to perform the canonical quantization of the
system. The variables $\Gamma^A$ and $\Pi_A$ are promoted to
operators and are required to satisfy the following equal time
commutation relations \be [\Gamma^A,\Pi_B]=i\b
h\delta^A\,_B.\label{commGP}\ee

The second class constraints $\Theta_A(\Gamma,\Pi)$ are converted
into Abelian first class by introducing the new bosonic operators
$\Phi_a$, $a=1,2,...,2N$ \cite{bf2}--\cite{bff2} which satisfy the
following equal time commutation relations \be
[\Phi_a,\Phi_b]=-i\b h\,\omega_{ab}^{(0)},\label{commPhi}\ee where
$\omega_{ab}^{(0)}$ is the same as in (\ref{sympmetric}). The
effective Abelian constraints are defined by the commutations
relations \be
[\T_A(\Gamma,\Pi,\Phi),\T_B(\Gamma,\Pi,\Phi)]=0\label{commT}\ee
and the boundary conditions
$\T_A(\Gamma,\Pi,0)=\Theta_A(\Gamma,\Pi)$.

We seek for the solution in the following form \be
\T_A(\Gamma,\Pi,\Phi)=\Theta_A(\Gamma,\Pi)+K_A(\Gamma,\Phi),
\qquad \qquad K_A(\Gamma,0)=0.\label{T}\ee

Substituting this expression into (\ref{commT}) we get the
following equation for $K_A(\Gamma,\Phi)$: \be
\nabla_AK_B-\nabla_BK_A+\Lambda^C\,_{AB}K_C-(i\b
h)^{-1}\left[K_A,K_B\right]=\omega_{AB}(\Gamma).\label{eqT}\ee

It is important here to note that the solution of (\ref{eqT}) in
the zero curvature case $\Lambda^C\,_{AB}=0$ is given by \be
K_A=\exp\left(\Phi_a\pa^a\right)\t K_A(\Gamma,\vp)|_{\vp=0},
\qquad \qquad \t K_A=\vp_a\,h^a\,_A(\Gamma),\label{K}\ee where
$\vp_a$ are classical variables and $\pa^a\equiv\pa/\pa\vp_a$.

We further construct from the initial first class constraints
$T'_i(\Gamma)$, $i=1,2,...,m'$ of our system, the quantities $\t
T'_i(\Gamma,\Phi)$ that commute with the effective Abelian
constraints $\T_A(\Gamma,\Pi,\Phi)$ \cite{bff}: \be [\t
T'_i(\Gamma,\Phi),\T_A(\Gamma,\Pi,\Phi)]=0, \quad
A=1,2,...,2N;\quad i=1,2,...,m';\label{commtTT}\ee with boundary
conditions $\t T'_i(\Gamma,0)=T'_i(\Gamma)$.

After substituting (\ref{T}) into (\ref{commtTT}) we obtain the
following equation for $\t T'_i(\Gamma,\Phi)$: \be \pa_A\t
T'_i(\Gamma,\Phi)=(i\b h)^{-1}\left[K_A(\Gamma,\Phi),\t
T'_i(\Gamma,\Phi)\right]. \label{deqtT'}\ee

It is worth noticing that the solution of the equation
(\ref{deqtT'}) in the special case $\Lambda^C\,_{AB}=0$ is given
by the following relation \cite{bf}: \be \t{
T}'_i(\Gamma,\Phi)=\exp\left(\Phi_a\pa^a\right)\t{\t{
T}}_i(\Gamma,\vp)|_{\vp=0},\label{soltT'}\ee where $\t{\t{
T}}_i(\Gamma,\vp)=T'_i\left(\overline\Gamma(x=0)\right)$.

The functions $\overline{\Gamma}^A(x)$ are solutions of the
differential equation \be\frac{d\overline{\Gamma}^A(x)}{dx}=
\vp_a\omega^{ab}_{(0)}h^A\,_b(\overline{\Gamma}),\qquad
\overline{\Gamma}^A(x=1)=\Gamma^A.\label{deGbar}\ee

Now we can construct the fermion generating operator $\Omega$. In
order to do that we introduce a pair (coordinate, conjugate
momentum) $(\lambda,\pi)$ together with a pair of ghosts
($C,\overline{\P}$) and antighosts ($P,\overline{\C}$) for every
irreducible first class constraint, where $\lambda$ is an active
Lagrange multiplier. The new introduced variables possess the
following statistics ($\eps$) and ghost number ($gh$) \cite{bff}:
\be
\eps(\lambda)=\eps(\pi)=1+\eps(C)=1+\eps(\overline{C})=1+\eps(\P)=1+\eps(\overline{\P})
\nonumber\ee \be
gh(C)=-gh(\overline{\P})=gh(\P)=-gh(\overline{C})=1.\label{statistics}\ee

Only the following supercommutators of the above introduced
operators are different from zero: \be
[\lambda,\pi]=[C,\overline{\P}]=[P,\overline{C}]=i\b h
I,\label{commlP}\ee where $I$ is the unit matrix with the
subindices labelling the complete set of irreducible first class
constraints and the supercommutator is defined as follows: \be
[A,B]=AB-BA(-1)^{\eps(A)\eps(B)}.\label{scomm}\ee

In the case when initial second class constraints are absent, the
fermion generating operator $\Omega$ is given in the form \be
\Omega=\Omega'(\Gamma,\Phi,C',\overline{\P}')+\T_A(\Gamma,\Pi,\Phi)C^A+\Pi'_i\P'^i+
\pi_A\P^A,\label{Omega}\ee where the Fermi operator $\Omega'$
obeys the following equations \be
[\Omega',\Omega']=[\Omega',\T_A]=0, \qquad
gh(\Omega')=1,\label{commOmega}\ee with the boundary condition \be
\Omega'(\Gamma,\Phi,C',0)=\t{
T}'_i(\Gamma,\Phi)C'^{i}.\nonumber\ee

By virtue of equations (\ref{commOmega}), the operator
(\ref{Omega}) turns out to be nilpotent: \be [\Omega,\Omega]=0.
\label{nilpotent}\ee

This is a very brief and incomplete exposition of the canonical
quantization of systems with curved phase space introduced by I.A.
Batalin, E.S. Fradkin and T.E. Fradkina. Along this line we are
going to proceed in treating the four--dimensional
Einstein--Maxwell Dilaton--Axion theory.

\section{The four--dimensional EMDA system}

The EMDA system arises as one of the simplest low--energy string
gravity models. It arises as the corresponding truncation of the
critical heterotic string theory (D=10, with 16 $U(1)$ vector
fields) reduced to four dimensions with no moduli fields excited
and just one non--vanishing $U(1)$ vector field (see, for
instance, \cite{gk}--\cite{ggk} and references therein). In the
Einstein frame it is described by the action
\begin{eqnarray}
S = \frac{1}{16\pi}\int d^4 x |^{(4)}g|^{\frac {1}{2}} \left \{ -R
+ 2\pa_\mu\phi\pa^\mu\phi +
\frac{1}{2}e^{4\phi}\pa_\mu\kappa\pa^\mu\kappa -
e^{-2\phi}F_{\mu\nu}F^{\mu\nu} - \kappa F_{\mu\nu}\t F^{\mu\nu}
\right \}, \label{emda}
\end{eqnarray}
where \be F_{\mu\nu}=\partial_{\mu}A_{\nu}-\partial_{\nu}A_{\mu},
\qquad\qquad \t
F^{\mu\nu}=\frac{1}{2}E^{\mu\nu\lambda\sigma}F_{\lambda\sigma},
\nonumber\ee are the strength of the $U(1)$ vector field $A_{\mu}$
and its dual tensor, respectively, $R$ is the scalar curvature of
the gravitationa field, $\phi$ is the dilaton field, $\kappa$ is
the pseudoscalar axion field, and $E^{\mu \nu \lambda
\sigma}=\sqrt{|g|}\eps^{\mu \nu \lambda \sigma}$. Formally, the
EMDA theory can be considered as an extension of the
Einstein--Maxwell system to the case when one takes into account
the (pseudo)scalar dilaton and axion fields.

Using the ADM decomposition of the four--dimensional metric tensor
\cite{adm}--\cite{cnp} we have \be ^{(4)}g_{\mu\nu}=\left(\ba{cc}
N^iN_i-N_{\perp}^2&N_j\cr N_i&g_{ij}\ea \right),\qquad
^{(4)}g^{\mu\nu}=\left(\ba{cc} -N_{\perp}^{-2} &
N^j/N_{\perp}^{2}\cr N^i/N_{\perp}^{2} &
g^{ij}-N^iN^j/N_{\perp}^{2}\ea \right).\label{admdecomp}\ee

Thus, the ADM action is given by \cite{adm}--\cite{hrt} \be
I_{Gr}=\int dtdx^3\left(\pi^{ij}\dot
g_{ij}-N^{\perp}\H_{\perp}-N^i\H_i\right),\label{admaction}\ee
where \be
\H_{\perp}=g^{-1/2}\left(\pi_{ij}\pi^{ij}-\frac{1}{2}\Pi^2\right)-g^{1/2}\
^{(3)}R, \label{Hperp}\ee \be \H_i=
-2g_{ik}\pi^{kj}\,_{,j}-2\left(g_{ik,j}-g_{kj,i}\right)\pi^{kj},
\label{Hi}\ee $g_{ij}$ is the three--dimensional spatial metric,
$g\equiv {\rm det}g_{ij}$ and $\Pi=g_{ij}\pi^{ij}$.

We consider the spherically symmetric case
\cite{bcmn}--\cite{bct}. In the basis $(dr,d\theta,d\vp)$ we have
\be g_{ij}=\mbox{{\rm
diag}}\left(e^{2\mu(r,t)},e^{2\lambda(r,t)},e^{2\lambda(r,t)}\sin^2\theta\right),
\label{gij}\ee so that the spatial line element has the form \be
g_{ij}dx^idx^j=
e^{2\mu(r,t)}dr^2+e^{2\lambda(r,t)}\left(d\theta^2+\sin^2\theta
d\vp^2\right). \label{spatialint}\ee

The conjugate momenta are given by the expressions \be
\pi^{ij}=\mbox{{\rm
diag}}\left(\frac{1}{2}\pi_\mu(r,t)e^{-2\mu(r,t)}\sin\theta,\ \,
\frac{1}{4}\pi_\lambda(r,t)e^{-2\lambda(r,t)}\sin\theta,\ \,
\frac{1}{4}\pi_\lambda(r,t)e^{-2\lambda(r,t)}(\sin\theta)^{-1}\right).
\label{momenta}\ee

After integrating over the angles, the action (\ref{admaction})
becomes \cite{bcmn}: \be I_{Gr}=4\pi\int
dtdr\left(\pi_\mu\dot\mu+\pi_\lambda\dot\lambda-N^{\perp}\H_{\perp}-N^r\H_r\right),
\label{actgrav}\ee where \be
\H_{\perp}=e^{-\mu-2\lambda}\left[\frac{1}{8}\left(\pi_\mu^2-2\pi_\mu\pi_\lambda\right)
+2e^{4\lambda}\left(2\lambda''+3\lambda'^2-2\lambda'\mu'-e^{2(\mu-\lambda)}\right)
\right], \label{Htransv}\ee \be
\H_r=\pi_\lambda\lambda'+\pi_\mu\mu'-\pi_\mu',\qquad
\H_\theta=\H_\vp=0, \label{Hr}\ee and the primes denote spatial
derivatives.

Apart from an overall factor $\frac{1}{16\pi}$ in the action
(\ref{emda}), the part of the Lagrangian density concerning the
dilaton can be written as an expression first order in time
derivatives of the dilaton field \be L_d\equiv 2|^{(4)}g|^{\frac
{1}{2}}
\pa_\mu\phi\pa^\mu\phi=\pi_\p\dot\p+\frac{N^{\perp}}{8g^{1/2}}\pi_\p^2-\pi_\p
N^i\pa_i\p+ 2N^{\perp}g^{1/2}g^{ij}\pa_i\p\pa_j\p,\label{Ld}\ee
where, as it was pointed out above, $g^{ij}$ is the
three--dimensional spatial metric in the ADM decomposition.

In the same way, for the axion field we obtain the following
relation \be L_k\equiv \frac{1}{2}|^{(4)}g|^{\frac {1}{2}}e^{4\p}
\pa_\mu\kappa\pa^\mu\kappa=\pi_\kappa\dot\kappa+\frac{N^{\perp}}{2g^{1/2}}e^{-4\p}\pi_\kappa^2-
\pi_\kappa
N^i\pa_i\kappa+\frac{1}{2}N^{\perp}e^{4\p}g^{1/2}g^{ij}\pa_i\kappa\pa_j\kappa,\label{Lk}\ee
and for the $U(1)$ vector field we get \be L_{U(1)}\equiv
-|^{(4)}g|^{\frac {1}{2}}\left(e^{-2\phi}F_{\mu\nu}F^{\mu\nu}\!+\!
\kappa F_{\mu\nu}\t F^{\mu\nu}\right)\!=\!\pi^i\dot
A_i\!-\!\left[\frac{(N^{\perp})^2}{2}e^{2\p}\pi_{U(1)}^k+2N^{\perp}g^{1/2}g^{lk}N^jF_{jl}
\right.+\nonumber\ee \be \left.2(N^{\perp})^3g^{1/2}\kappa
e^{2\p}\t F^{0k}\right]
\left[\frac{1}{4N^{\perp}g^{1/2}}{\pi_{U(1)}}_k+e^{-2\p}\frac{N^j}{(N^{\perp})^2}F_{jk}+
\kappa g_{ik}\t F^{0i}\right]- \label{Lu} \ee\be
N^{\perp}g^{1/2}e^{-2\p}
\left(g^{ij}g^{kl}-2g^{ij}\frac{N^kN^l}{(N^{\perp})^2}\right)F_{jl}F_{ik}-A_0\pa_i\pi^i.
\nonumber\ee

Now we restrict ourselves to the spherically symmetric case and,
moreover, wee assume manifest spherical symmetry for the fields
\cite{bct} which means that the scalar fields depend only on $r$,
$t$ and that the vector fields have only the radial component that
depend on $r$, $t$. We define as well $p_\p(r,t)$,
$p_\kappa(r,t)$, $\E^r(r,t)$ as follows \be
\pi_\p=p_\p(r,t)\sin\theta,\qquad
\pi_\kappa=e^{4\p}p_\kappa(r,t)\sin\theta,\qquad
\pi^r=e^{-2\p}\E^r(r,t)\sin\theta.\label{ps}\ee

Finally, after integrating over the angles, the EMDA action
(\ref{emda}) takes the form \be
I=\frac{1}{4}\int\L_{tot}dtdr,\label{emdasph}\ee where
\be\L_{tot}=-\pi_\mu\dot\mu-\pi_\lambda\dot\lambda+\pi_\p\dot\p+e^{4\p}p_\kappa\dot\kappa+
e^{-2\p}\E^r\dot A_r+ \nonumber\ee \be
N^{\perp}\left\{e^{-\mu-2\lambda}\left[\frac{1}{8}\left(\pi_\mu^2-2\pi_\mu\pi_\lambda\right)
+2e^{4\lambda}\left(2\lambda''+3\lambda'^2-2\lambda'\mu'-e^{2(\mu-\lambda)}\right)
\right]+\right.\label{Ltot}\ee \be
\left.\frac{p_\p^2}{8}e^{-(\mu+2\lambda)}+2\p'^2e^{2\lambda-\mu}+\frac{p_\kappa^2}{2}
e^{4\p-\mu-2\lambda}+\frac{\kappa'^2}{2}e^{4\p-\mu+2\lambda}-
\frac{(\E^r)^2}{8}e^{-2\p+\mu-2\lambda}\right\}-\nonumber\ee \be
N^1\left(\pi_\mu'
-\pi_\mu\mu'-\pi_\lambda\lambda'+p_p\p'+e^{4\p}p_\kappa\kappa'\right)-
A_0e^{-2\p}\left({\E^r}'-2\p'\E^r\right).\nonumber\ee

Now we proceed as in \cite{bcmn} by imposing the coordinate
condition \be r=e^\lambda\label{coordcond}\ee and solving for
$\pi_\lambda$ the equation which results after putting the second
constraint (the one multiplied by $N^1$) equal to zero. Then
(\ref{Ltot}) is modified as follows
\be\L_{tot}=-\pi_\mu\dot\mu+\pi_\p\dot\p+e^{4\p}p_\kappa\dot\kappa+e^{-2\p}\E^r\dot
A_r+ \nonumber\ee \be
N^{\perp}e^{-\mu}\left\{\frac{1}{r^2}\left[\frac{1}{8}\pi_\mu^2-
\frac{r}{4}\pi_\mu\left(\pi_\mu'-\pi_\mu\mu'+p_\p\p'+e^{4\p}p_\kappa\kappa'\right)
\right]+\right.\label{Ltotcc}\ee \be
\left.2\!-\!4r\mu'\!-\!2e^{2\mu}+
\frac{p_\p^2}{8r^2}+2r^2\p'^2+e^{4\p}\frac{p_\kappa^2}{2r^2}+
e^{4\p}r^2\frac{\kappa'^2}{2}\!-\!e^{2(\mu-\p)}
\frac{(\E^r)^2}{8r^2}\right\}-A_0e^{-2\p}\left({\E^r}'-2\p'\E^r\right).
\nonumber\ee We se that (\ref{Ltotcc}) has eight variables
spanning a curved phase space and two constraints. Now we can
apply the I.A. Batalin, E.S. Fradkin and T.E. Fradkina canonical
quantization to this model.

\section{Canonical quantization of the EMDA system}

In order to be compatible with the notation of the Introduction we
rename the field variables of (\ref{Ltotcc}) in the following way
\be p_\p=\Gamma^1, \qquad \E^r=\Gamma^2, \qquad p_\kappa=\Gamma^3,
\qquad \pi_{\mu}=\Gamma^4, \nonumber\ee \be \p=\Gamma^5, \qquad
A_r=\Gamma^6, \qquad \kappa=\Gamma^7, \qquad \mu=\Gamma^8.
\label{Gs}\ee

Then , the ``canonical one--form" of (\ref{Ltotcc}) can be written
as follows up to a total time derivative \be
\pi_\p\dot\p+e^{4\p}p_\kappa\dot\kappa+e^{-2\p}\E^r\dot
A_r-\pi_\mu\dot\mu=a_A(\Gamma)\dot\Gamma^A, \qquad
A=1,2,...,8;\label{1form}\ee where \be a_1=-\frac{1}{2}\Gamma^5,
\qquad a_2=-\frac{1}{2}e^{-2\Gamma^5}\Gamma^6, \qquad
a_3=-\frac{1}{2}e^{4\Gamma^5}\Gamma^7, \qquad
a_4=\frac{1}{2}\Gamma^8, \nonumber\ee\be
a_5=\frac{1}{2}\left(\Gamma^2\!-\!4e^{4\Gamma^5}\Gamma^3\Gamma^7\!+\!
2e^{-2\Gamma^5}\Gamma^2\Gamma^6\right), \quad
a_6=\frac{1}{2}e^{-2\Gamma^5}\Gamma^2, \quad
a_7=\frac{1}{2}e^{4\Gamma^5}\Gamma^3, \quad
a_8=-\frac{1}{2}\Gamma^4. \label{aes}\ee

The symplectic metric is given by the relation \be
\omega_{AB}=\frac{\pa a_B}{\pa\Gamma^A}-\frac{\pa
a_A}{\pa\Gamma^B}.\label{wa}\ee

This is an invertible matrix which possesses the following form
\be \omega_{AB}=\left(\ba{cccccccc} 0 & 0 & 0 & 0 & 1 & 0 & 0 & 0
\cr 0 & 0 & 0 & 0 & 0 & e^{-2\Gamma^5} & 0 & 0 \cr 0 & 0 & 0 & 0 &
0 & 0 & e^{4\Gamma^5} & 0 \cr 0 & 0 & 0 & 0 & 0 & 0 & 0 & -1 \cr
-1 & 0 & 0 & 0 & 0 & -2e^{-2\Gamma^5}\Gamma^2 &
4e^{4\Gamma^5}\Gamma^3 & 0 \cr 0 & -e^{-2\Gamma^5} & 0 & 0 &
2e^{-2\Gamma^5}\Gamma^2 & 0 & 0 & 0 \cr 0 & 0 & -e^{4\Gamma^5} & 0
& -4e^{4\Gamma^5}\Gamma^3 & 0 & 0 & 0 \cr 0 & 0 & 0 & 1 & 0 & 0 &
0 & 0 \ea\right).\label{matrixw}\ee

The expression for the inverse matrix reads \be
\omega^{AB}=\left(\ba{cccccccc} 0 & -2\Gamma^2 & 4\Gamma^3 & 0 &
-1 & 0 & 0 & 0 \cr 2\Gamma^2 & 0 & 0 & 0 & 0 & -e^{2\Gamma^5} & 0
& 0 \cr -4\Gamma^3 & 0 & 0 & 0 & 0 & 0 & -e^{-4\Gamma^5} & 0 \cr 0
& 0 & 0 & 0 & 0 & 0 & 0 & 1 \cr 1 & 0 & 0 & 0 & 0 & 0 & 0 & 0 \cr
0 & e^{2\Gamma^5} & 0 & 0 & 0 & 0 & 0 & 0 \cr 0 & 0 &
e^{-4\Gamma^5} & 0 & 0 & 0 & 0 & 0 \cr 0 & 0 & 0 & -1 & 0 & 0 & 0
& 0 \ea\right).\label{matrixinvw}\ee

We define as well the equal time Poisson bracket for any two
functions $X\left(\Gamma(r,t),\Gamma'(r,t)\right)$ and
$Y\left(\Gamma(r,t),\Gamma'(r,t)\right)$ as follows \be
\left\{X\left(\Gamma(r,t),\Gamma'(r,t)\right),
Y\left(\Gamma(r',t),\Gamma'(r',t)\right)\right\}=\nonumber\ee\be
\int
dr''\left[\frac{\pa}{\pa\Gamma^A(r'',t)}X\left(\Gamma,\Gamma'\right)
\omega^{AB}\left(\Gamma(r'',t)\right)
\frac{\pa}{\pa\Gamma^B(r'',t)}Y\left(\Gamma,\Gamma'\right)\right],\label{XYrr'}\ee
where the prime over the $\Gamma$ denotes derivative with respect
to the spatial variable.

There are two irreducible constraints in the action
(\ref{Ltotcc}), namely \be T'_1\left(\Gamma,\Gamma'\right)=
e^{-\Gamma^8}\left\{\frac{1}{8r^2}\left[\left(\Gamma^4\right)^2-
2r\Gamma^4\left(\Gamma^{4'}-\Gamma^4\Gamma^{8'}+\Gamma^1\Gamma^{5'}+
e^{4\Gamma^5}\Gamma^3\Gamma^{7'}\right)\right]+2-4r\Gamma^{8'}-\right.
\nonumber\ee\be \left.2e^{2\Gamma^8}+
\frac{\left(\Gamma^1\right)^2}{8r^2}+2r^2\left(\Gamma^{5'}\right)^2+
e^{4\Gamma^5}\frac{\left(\Gamma^3\right)^2}{2r^2}+
e^{4\Gamma^5}r^2\frac{\left(\Gamma^{7'}\right)^2}{2}-
e^{2\left(\Gamma^8-\Gamma^5\right)}\frac{\left(\Gamma^2\right)^2}{8r^2}
\right\} \label{T1'}\ee and \be T'_2\left(\Gamma,\Gamma'\right)=
e^{-2\Gamma^5}\left(2{\Gamma^2\Gamma^5}'-{\Gamma^2}'\right).\label{T2'}\ee

These  are first class constraints as they should be
\cite{dirac1}--\cite{schwinger} (see \cite{hrt} as well) and their
Poisson brackets satisfy the following relations: \be
\left\{T'_1\left(\Gamma(r,t),\Gamma'(r,t)\right),
T'_1\left(\Gamma(r',t),\Gamma'(r',t)\right)\right\}=\nonumber\ee\be
\frac{\Gamma^4(r,t)}{4r}T'_1\left(\Gamma(r,t),\Gamma'(r,t)\right)\frac{\pa}{\pa
r'}\delta(r'-r)-(r\longleftrightarrow r'),\label{T1T1}\ee \be
\left\{T'_1\left(\Gamma(r,t),\Gamma'(r,t)\right),
T'_2\left(\Gamma(r',t),\Gamma'(r',t)\right)\right\}=0,\label{T1T2}\ee
\be \left\{T'_2\left(\Gamma(r,t),\Gamma'(r,t)\right),
T'_2\left(\Gamma(r',t),\Gamma'(r',t)\right)\right\}=0.\label{T2T2}\ee

The next step is to determine the reper field $h^a\,_A$ from the
relations (\ref{haA}) in such a way that the curvature
$\Lambda^C_{AB}$ defined in the relation (\ref{Lambda}) is zero.
This is a very crucial point because it simplifies enormously the
forthcoming calculations. One such solution is given by the
following matrix \be h^a\,_A(\Gamma)= \left(\ba{cccccccc} 1 & 0 &
0 & 0 & 0 & 0 & 0 & 0 \cr 0 & e^{-2\Gamma^5} & 0 & 0 & 0 & 0 & 0 &
0 \cr 0 & 0 & e^{4\Gamma^5} & 0 & 0 & 0 & 0 & 0 \cr 0 & 0 & 0 & -1
& 0 & 0 & 0 & 0 \cr 0 & -2e^{-2\Gamma^5}\Gamma^2 &
4e^{4\Gamma^5}\Gamma^3 & 0 & 1 & 0 & 0 & 0 \cr 0 & 0 & 0 & 0 & 0 &
1 & 0 & 0 \cr 0 & 0 & 0 & 0 & 0 & 0 & 1 & 0 \cr 0 & 0 & 0 & 0 & 0
& 0 & 0 & 1 \ea\right).\label{matrixhaA}\ee Its inverse matrix has
the following form \be h^A\,_a(\Gamma)= \left(\ba{cccccccc} 1 & 0
& 0 & 0 & 0 & 0 & 0 & 0 \cr 0 & e^{2\Gamma^5} & 0 & 0 & 0 & 0 & 0
& 0 \cr 0 & 0 & e^{-4\Gamma^5} & 0 & 0 & 0 & 0 & 0 \cr 0 & 0 & 0 &
-1 & 0 & 0 & 0 & 0 \cr 0 & 2\Gamma^2 & -4\Gamma^3 & 0 & 1 & 0 & 0
& 0 \cr 0 & 0 & 0 & 0 & 0 & 1 & 0 & 0 \cr 0 & 0 & 0 & 0 & 0 & 0 &
1 & 0 \cr 0 & 0 & 0 & 0 & 0 & 0 & 0 & 1
\ea\right),\label{matrixhAa}\ee and \be \omega_{ab}^{(0)}=
\left(\ba{cccccccc} 0 & 0 & 0 & 0 & 1 & 0 & 0 & 0 \cr 0 & 0 & 0 &
0 & 0 & 1 & 0 & 0 \cr 0 & 0 & 0 & 0 & 0 & 0 & 1 & 0 \cr 0 & 0 & 0
& 0 & 0 & 0 & 0 & 1 \cr -1 & 0 & 0 & 0 & 0 & 0 & 0 & 0 \cr 0 & -1
& 0 & 0 & 0 & 0 & 0 & 0 \cr 0 & 0 & -1 & 0 & 0 & 0 & 0 & 0 \cr 0 &
0 & 0 & -1 & 0 & 0 & 0 & 0 \ea\right).\label{w0}\ee

It is worth noticing that with this choice of the reper field
components $h^a\,_A=0$ we have $\Lambda^C_{AB}=0$, a fact which
will greatly simplify the computations when deriving the explicit
expressions for some quantities. We have checked as well that the
covariant derivative of the symplectic metric (\ref{matrixw}) and
the reper field components (\ref{matrixhaA}) vanish, i.e.,
$\nabla_C\omega_{AB}=0$ and $\nabla_C\,h^a\,_A=0$.

Now we introduce new bosonic fields $\Pi_A(r,t)$, which we promote
into operators along with the variables $\Gamma^A(r,t)$. We also
define the nonzero commutators as follows: \be
[\Gamma^A(r,t),\Pi_B(r',t)]=i\b
h\delta^A\,_B\delta(r-r').\label{commGPrr'}\ee

This is a slightly modified version of the relation (\ref{commGP})
since it involves the Dirac $\delta$--function which is not
present in its original definition.

The special second class constraints $\Theta_A(\Gamma,\Pi)$,
defined in (\ref{Thetas}), are given by the following expression
\be \Theta _1(r,t)=\Pi_1(r,t)+\frac{1}{2}\Gamma^5(r,t), \qquad
\Theta
_2(r,t)=\Pi_2(r,t)+\frac{1-e^{-2\Gamma^5(r,t)}\left(1+2\Gamma^5(r,t)\right)}
{4\left(\Gamma^5(r,t)\right)^2}\Gamma^6(r,t),\nonumber\ee \be
\Theta
_3(r,t)=\Pi_3(r,t)+\frac{1+e^{4\Gamma^5(r,t)}\left(-1+4\Gamma^5(r,t)\right)}
{16\left(\Gamma^5(r,t)\right)^2}\Gamma^7(r,t),\qquad \Theta
_4(r,t)=\Pi_4(r,t)-\frac{1}{2}\Gamma^8(r,t), \nonumber\ee \be
\Theta _5(r,t)=\Pi_5(r,t)-\frac{1}{2}\Gamma^1(r,t)-
\frac{1-e^{-2\Gamma^5(r,t)}\left(1+2\Gamma^5(r,t)+2\left(\Gamma^5(r,t)\right)^2\right)}
{2\left(\Gamma^5(r,t)\right)^3}\Gamma^2(r,t)\Gamma^6(r,t)-\nonumber\ee
\be
\frac{1-e^{4\Gamma^5(r,t)}\left(1-4\Gamma^5(r,t)+8\left(\Gamma^5(r,t)\right)^2\right)}
{8\left(\Gamma^5(r,t)\right)^3}\Gamma^3(r,t)\Gamma^7(r,t),\label{Thetasexplicit}\ee
\be \Theta _6(r,t)=\Pi_6(r,t)
+\frac{1-e^{-2\Gamma^5(r,t)}\left(1+2\Gamma^5(r,t)+4\left(\Gamma^5(r,t)\right)^2\right)}
{4\left(\Gamma^5(r,t)\right)^2}\Gamma^2(r,t),\nonumber\ee \be
\Theta _7(r,t)=\Pi_7(r,t)
+\frac{1-e^{4\Gamma^5(r,t)}\left(1-4\Gamma^5(r,t)+16\left(\Gamma^5(r,t)\right)^2\right)}
{16\left(\Gamma^5(r,t)\right)^2}\Gamma^3(r,t), \nonumber\ee \be
\Theta _8(r,t)=\Pi_8(r,t)+\frac{1}{2}\Gamma^4(r,t). \nonumber\ee

These quantities satisfy the following commutation relations
\be\left[\Theta_A(r,t),\Theta_B(r',t)\right]=i\b
h\,\omega_{AB}(r,t)\delta(r-r')\label{Thetasrr'}\ee as well as the
following flat Poisson bracket relations
\be\{\Theta_A(r,t),\Theta_B(r',t)\}'\equiv \int
dr''\left[\pa_C\Theta_A(r,t)\ \pa^C\Theta_B(r',t)-
\pa_C\Theta_B(r',t)\ \pa^C\Theta_A(r,t)\right]=\nonumber\ee\be
\omega_{AB}(r,t)\delta(r-r'),\label{Thetas'rr'}\ee where \be
\pa_C\equiv\frac{\pa}{\pa\Gamma^C(r'',t)} \qquad \mbox{{\rm and}}
\qquad \pa^C\equiv\frac{\pa}{\pa\Pi_C(r'',t)}.\nonumber\ee

Next, as prescribed in the Introduction, the second class
constraints $\Theta_A$ are converted into Abelian first class
constraints by the introduction of the new bosonic fields
$\Phi_a$, $a=1,2,...,8$ which satisfy the following equal time
commutation relations \be [\Phi_a(r,t),\Phi_b(r',t)]=-i\b
h\,\omega_{ab}^{(0)}\delta(r-r'),\label{commPhirr'}\ee where
$\omega_{ab}^{(0)}$ is defined as in (\ref{w0}). We are
considering the case $\Lambda^C_{AB}=0$, thus, from (\ref{T}) and
(\ref{eqT}) the first class constraints $\T_A(\Gamma,\Pi,\Phi)$
are given by \be
\T_A(\Gamma,\Pi,\Phi)=\Theta_A(\Gamma,\Pi)+\K_A(\Gamma,\Phi)=
\Theta_A(\Gamma,\Pi)+\Phi_a\ h^a\,_A(\Gamma). \label{Texplicit}\ee

Indeed, we can show that these first class constraints
$\T_A(\Gamma,\Pi,\Phi)$ are Abelian since \be
\left[\T_A\left(r,t\right), \T_B\left(r',t\right)\right]=
\left[\Theta_A(r,t)+\Phi_a(r,t)h^a\,_A(r,t),
\Theta_B(r',t)+\Phi_b(r',t)h^b\,_B(r',t)\right]=\nonumber\ee \be
\left[\Theta_A(r,t),\Theta_B(r',t)\right]+
\left[\Phi_a(r,t)h^a\,_A(r,t),\Phi_b(r',t)h^b\,_B(r',t)\right]=\label{commTATB}\ee
\be i\b h\,\omega_{AB}\delta(r-r')-i\b
h\,h^a\,_A(\Gamma)\omega_{ab}^{(0)}h^b\,_B(\Gamma)\delta(r-r')=0.\nonumber\ee
Now we proceed to construct the operators $\t
T'_1(\Gamma,\Gamma',\Phi,\Phi')$ and $\t
T'_2(\Gamma,\Gamma',\Phi,\Phi')$ so that their commutator with
$\T_A$ vanishes \be \left[\t
T'_i(\Gamma,\Gamma',\Phi,\Phi'),\T_A(\Gamma,\Pi,\Phi)\right]=0
\label{commtT'TA}\ee and \be \t T'_i(\Gamma,\Gamma',0,0)=
T'_i(\Gamma,\Gamma'), \qquad  i=1,2;\nonumber\ee where $
T'_i(\Gamma,\Gamma')$ are the first class constraints given in
(\ref{T1'}) and (\ref{T2'}). Here we have to note that the
quantities $\t T'_i$ depend note only on the $\Phi_a$ fields, but
also on their space derivatives and this is because of the
existence of space derivatives of the phase space variables
$\Gamma^A$ in the initial first class constraints $T'_1$ and
$T'_2$. By substituting the relation (\ref{Texplicit}) into
(\ref{commtT'TA}) we obtain \be \left[\t
T'_i(\Gamma(r,t),\Gamma'(r,t),\Phi(r,t),\Phi'(r,t)),\
\,\Pi_A(r',t) \right]+\nonumber\ee \be \left[\t
T'_i(\Gamma(r,t),\Gamma'(r,t),\Phi(r,t),\Phi'(r,t)),\
\,\Phi_a(r',t)h^a\,_A(r',t)\right]=0. \label{ctT'Fi}\ee Due to the
relation (\ref{commGPrr'}) this equation adopts the following form
\be \frac{\delta}{\delta\Gamma^A(r',t)}\t
T'_i(\Gamma(r,t),\Gamma'(r,t),\Phi(r,t),\Phi'(r,t))= \nonumber\ee
\be \left(i\b h\right)^{-1} \left[\Phi_a(r',t)h^a\,_A(r',t),\ \,\t
T'_i(\Gamma(r,t),\Gamma'(r,t),\Phi(r,t),\Phi'(r,t))\right].
\label{FitT'}\ee

Let us try to find a solution for equation (\ref{FitT'}) in the
case when the constraint $T'_2\left(\Gamma,\Gamma'\right)$ is
given by the relation (\ref{T2'}) \be
T'_2\left(\Gamma,\Gamma'\right)=
e^{-2\Gamma^5}\left(2{\Gamma^2\Gamma^5}'-{\Gamma^2}'\right).\nonumber\ee

We shall consider the function $T_2\left(\Gamma\right)=
-e^{-2\Gamma^5}\left(\Gamma^2-2\Gamma^2\Gamma^5\right)$. This
relation comes from the expression of
$T'_2\left(\Gamma,\Gamma'\right)$ after ignoring the primes. Now
we turn to solve the equation \be
\frac{\delta}{\delta\Gamma^A(r',t)}\t T_2(\Gamma(r,t),\Phi(r,t))=
\left(i\b h\right)^{-1} \left[\Phi_a(r',t)h^a\,_A(r',t),\ \,\t
T_2(\Gamma(r,t),\Phi(r,t))\right] \label{FitT2}\ee with \be \t
T_2(\Gamma(r,t),0)=T_2(\Gamma(r,t)). \nonumber\ee The solution of
equation (\ref{FitT2}) is given in the form \be \t
T_2(\Gamma(r,t),\Phi(r,t))= \exp\left(\int
dr'\Phi_a(r',t)\frac{\pa}{\pa\Phi_a(r',t)}\right)
T_2(\overline\Gamma(x=0))\left|_{\vp=0}\right., \label{tT2sol}\ee
where the functions $\overline\Gamma^A(x)$ are solutions of the
equations (\ref{deGbar}) and are given by the following relations
\be \overline\Gamma^1(x)= \vp_5(x-1)+\Gamma^1, \qquad
\overline\Gamma^2(x)=e^{-2\vp_1(x-1)}\left[\Gamma^2+\vp_6e^{2\Gamma^5}(x-1)\right],
\nonumber\ee \be
\overline\Gamma^3(x)=e^{4\vp_1(x-1)}\left[\Gamma^3+\vp_7e^{-4\Gamma^5}(x-1)\right],
\qquad \overline\Gamma^4(x)= \vp_8(1-x)+\Gamma^4, \nonumber\ee \be
\overline\Gamma^5(x)= \vp_1(1-x)+\Gamma^5, \qquad
\overline\Gamma^6(x)= \vp_2(1-x)+\Gamma^6, \label{Gbar}\ee \be
\overline\Gamma^7(x)= \vp_3(1-x)+\Gamma^7, \qquad
\overline\Gamma^8(x)= \vp_4(1-x)+\Gamma^8. \nonumber\ee

Thus, from equation (\ref{tT2sol}) we obtain for the solution $\t
T_2\left(\Gamma(r,t),\Phi(r,t)\right)$ of equation (\ref{FitT2})
the following expression \be \t T_2(\Gamma(r,t),\Phi(r,t))=-
e^{-2\left(\Phi_1+\Gamma^5\right)}\left[
e^{2\Phi_1}\left(\Gamma^2\!-\!\Phi_6e^{2\Gamma^5}\right)\!-\!2\left(\Phi_1+\Gamma^5\right)
e^{2\Phi_1}\left(\Gamma^2\!-\!\Phi_6e^{2\Gamma^5}\right)\right].
\label{tT2=}\ee

From the relation (\ref{tT2=}) we can write down a solution of
(\ref{FitT'}) empirically in the form \be \t
T'_2\left(\Gamma(r,t),\Gamma'(r,t),\Phi(r,t),\Phi'(r,t)\right)=
-e^{-2\left(\Phi_1+\Gamma^5\right)}\left\{\frac{d}{dr}
\left[e^{2\Phi_1}\left(\Gamma^2-\Phi_6e^{2\Gamma^5}\right)\right]\right.-\nonumber\ee
\be
2\left.\left(\Phi_1'+{\Gamma^5}'\right)e^{2\Phi_1}\left(\Gamma^2-\Phi_6e^{2\Gamma^5}\right)
\right\}. \label{tT2'}\ee

It is a remarkable fact that we can establish an analogy between
each term of the constraint $\t
T'_2\left(\Gamma(r,t),\Gamma'(r,t),\Phi(r,t),\Phi'(r,t)\right)$ of
the equation (\ref{tT2'}) and each term of the constraint
$T'_2(\Gamma,\Gamma')$ of the equation (\ref{T2'}). The simplified
expression for the equation (\ref{tT2'}) reads \be \t
T'_2(\Gamma,\Gamma',\Phi,\Phi')= -e^{-2\Gamma^5}\left(
{\Gamma^2}'-\Phi_6'e^{2\Gamma^5}- 2{\Gamma^5}'\Gamma^2\right).
\label{tT2'simp}\ee

Let us check now that the expression (\ref{tT2'simp}) satisfies
the equation (\ref{FitT'}). For instance we can compute the
following relationship \be \frac{\delta}{\delta\Gamma^2(r',t)}\t
T_2\left(\Gamma(r,t),\Gamma'(r,t),\Phi(r,t),\Phi'(r,t)\right)=\!-e^{-2\Gamma^5(r,t)}
\!\left[\frac{\pa}{\pa
r}\delta(r\!-\!r')\!-\!2{\Gamma^5}'(r,t)\delta(r\!-\!r')\right]\!.
\label{dtT2G2}\ee On the other hand we have \be
\left[\Phi_2(r',t)h^2\,_2(r',t),\ \,\t
T'_2\left(\Gamma(r,t),\Gamma'(r,t),\Phi(r,t),\Phi'(r,t)\right)\right]=e^{-2\Gamma^5(r',t)}
\left[\Phi_2(r',t),\Phi_6'(r,t)\right]=\nonumber\ee \be -i\b h
e^{-2\Gamma^5(r',t)}\frac{\pa}{\pa r}\delta(r-r')=-i\b h
e^{-2\Gamma^5(r,t)}\left[\frac{\pa}{\pa
r}\delta(r-r')-2{\Gamma^5}'(r,t)\delta(r-r')\right],
\label{cFi2tT2'}\ee where we have made use of the relations
(\ref{A1}) and (\ref{A2}) of the Appendix A. It is easy to see
that (\ref{tT2'simp}) satisfies the equation (\ref{FitT'}) in all
cases and also that the boundary condition $\t
T_2'(\Gamma,\Gamma',0,0)=T'_2(\Gamma,\Gamma')$ is satisfied.

We turn now to solve the equation (\ref{FitT'}) for the case when
the constraint $T'_1(\Gamma,\Gamma')$ defined in (\ref{T1'}). In
order to do that we shall follow the same procedure as we did with
the constraint $T'_2(\Gamma,\Gamma')$. Let us consider the
function \be
T_1\left(\Gamma\right)=e^{-\Gamma^8}\left\{\frac{1}{8r^2}
\left[\left(\Gamma^4\right)^2-2r\Gamma^4
\left(\Gamma^4-\Gamma^4\Gamma^8+\Gamma^1\Gamma^5+
e^{4\Gamma^5}\Gamma^3\Gamma^7\right)\right]+2-4r\Gamma^8-2e^{2\Gamma^8}+\right.\nonumber\ee
\be
\left.\frac{\left(\Gamma^1\right)^2}{8r^2}+2r^2\left(\Gamma^5\right)^2+
e^{4\Gamma^5}\frac{\left(\Gamma^3\right)^2}{2r^2}+
\frac{r^2}{2}e^{4\Gamma^5}\left(\Gamma^7\right)^2-
e^{2\left(\Gamma^8-\Gamma^5\right)}\frac{\left(\Gamma^2\right)^2
}{8r^2}\right\}. \label{T1}\ee

As in the previous case, the expression for $T_1(\Gamma)$ does
come from the constraint $T_1'(\Gamma,\Gamma')$ when ignoring the
primes. Thus, in order to solve the following equation \be
\frac{\delta}{\delta\Gamma^A(r',t)}\t T_1(\Gamma(r,t),\Phi(r,t))=
\left(i\b h\right)^{-1} \left[\Phi_a(r',t)h^a\,_A(r',t),\t
T_1(\Gamma(r,t),\Phi(r,t))\right] \label{FitT1}\ee with \be \t
T_1(\Gamma(r,t),0)=T_1(\Gamma(r,t)), \nonumber\ee we need the
expression for $T_1(\overline\Gamma(x=0))$ which is given by \be
T_1\left(\overline\Gamma(x=0)\right)=
e^{-\left(\vp_4+\Gamma^8\right)}\left\{\frac{1}{8r^2}
\left[\left(\vp_8+\Gamma^4\right)^2-2r\left(\vp_8+\Gamma^4\right)
\left[\left(\vp_8+\Gamma^4\right)- \right.\right.\right.
\nonumber\ee \be \left.\left.
\left(\vp_8+\Gamma^4\right)\left(\vp_4+\Gamma^8\right)\!+\!
\left(-\vp_5+\Gamma^1\right)\left(\vp_1+\Gamma^5\right)\!+\!
e^{4\Gamma^5}
\left(\Gamma^3-\vp_7e^{-4\Gamma^5}\right)\left(\vp_3+\Gamma^7\right)\right]\right]+
\nonumber\ee \be
2-4r\left(\vp_4+\Gamma^8\right)-2e^{2\left(\vp_4+\Gamma^8\right)}+
\frac{\left(-\vp_5+\Gamma^1\right)^2}{8r^2}+2r^2\left(\vp_1+\Gamma^5\right)^2+
\label{T1Gb} \ee \be
\left.\frac{e^{4\left(-\vp_1+\Gamma^5\right)}}{2r^2}
\left(\Gamma^3-\vp_7e^{-4\Gamma^5}\right)^2+e^{4\left(\vp_1+\Gamma^5\right)}
\frac{r^2}{2}\left(\vp_3+\Gamma^7\right)^2-
e^{2\left(\vp_4+\vp_1+\Gamma^8-\Gamma^5\right)}
\frac{\left(\Gamma^2-\vp_6e^{4\Gamma^5}\right)^2}{8r^2}\right\}.
\nonumber\ee

We are looking for a solution of the equation (\ref{FitT1}) in the
following form \be \t T_1(\Gamma(r,t),\Phi(r,t))= \exp\left(\int
dr'\Phi_a(r',t)\frac{\pa}{\pa\Phi_a(r',t)}\right)
T_1(\overline\Gamma(x=0))\left|_{\vp=0}\right.. \label{tT1sol}\ee

There is one technical problem here: not all of the $\Phi_a$
fields commute. We shall avoid this difficulty by making use of
the M. Suzuki's formula \cite{suzuki}, that we quote in
(\ref{A3}), in order to put these fields in a symmetric ordering.
Thus, we obtain the following result: \be \t
T_1\left(\Gamma(r,t),\Phi(r,t)\right)= \sum_{i=1}^8\t
T_{1i}\left(\Gamma(r,t),\Phi(r,t)\right),\label{tT1i} \ee where
\be \t T_{11}\left(\Gamma,\Phi\right)= \frac{1-2r}{8r^2}\left[
\left(\Gamma^4\right)^2e^{-\left(\Phi_4+\Gamma^8\right)}
+\frac{1}{2}\Phi_8e^{-\left(\Phi_4+\Gamma^8\right)}\Phi_8\right.+
\nonumber\ee \be
\left.\Gamma^4\left(\Phi_8e^{-\left(\Phi_4+\Gamma^8\right)}+
e^{-\left(\Phi_4+\Gamma^8\right)}\Phi_8\right)+\frac{1}{4}
\left(\Phi_8^2e^{-\left(\Phi_4+\Gamma^8\right)}+
e^{-\left(\Phi_4+\Gamma^8\right)}\Phi_8^2\right)\right],
\nonumber\ee \be \t T_{12}\left(\Gamma,\Phi\right)=
\frac{1}{4r}\left[
\left(\Gamma^4\right)^2C+\Gamma^4\left(\Phi_8C+C\Phi_8\right)+\frac{1}{2}
\left(\Phi_8C\Phi_8+ \Phi_8^2C+C\Phi_8^2\right)\right],
\nonumber\ee where
$C=\left(\Phi_4+\Gamma^8\right)e^{-\left(\Phi_4+\Gamma^8\right)}$;
\be \t T_{13}\left(\Gamma,\Phi\right)=-\frac{1}{4r}\left[
\Gamma^4e^{-\left(\Phi_4+\Gamma^8\right)}
+\frac{1}{2}\left(e^{-\left(\Phi_4+\Gamma^8\right)}\Phi_8+
\Phi_8e^{-\left(\Phi_4+\Gamma^8\right)}\right)\right]\times
\nonumber\ee \be
\left[\Gamma^1\Gamma^5+\left(\Gamma^1\Phi_1-\Gamma^5\Phi_5\right)-
\frac{1}{2}\left(\Phi_1\Phi_5+\Phi_5\Phi_1\right)\right],
\nonumber\ee \be \t
T_{14}\left(\Gamma,\Phi\right)=-\frac{1}{4r}e^{4\Gamma^5}
\left[\Gamma^4e^{-\left(\Phi_4+\Gamma^8\right)}
+\frac{1}{2}\left(e^{-\left(\Phi_4+\Gamma^8\right)}\Phi_8+
\Phi_8e^{-\left(\Phi_4+\Gamma^8\right)}\right)\right]\times
\nonumber\ee \be
\left[\Gamma^3\Gamma^7+\left(\Gamma^3\Phi_3-e^{-4\Gamma^5}
\Gamma^7\Phi_7\right)-
\frac{1}{2}e^{-4\Gamma^5}\left(\Phi_7\Phi_3+\Phi_3\Phi_7\right)\right],
\nonumber\ee \be \t
T_{15}\left(\Gamma,\Phi\right)=e^{-\left(\Phi_4+\Gamma^8\right)}
\left[2-2e^{2\left(\Phi_4+\Gamma^8\right)}
+\frac{1}{8r^2}\left(\Phi_5-\Gamma^1\right)^2\right.- \nonumber\ee
\be \left. \frac{1}{8r^2}
e^{2\left(\Phi_4+\Gamma^8\right)}e^{2\left(\Phi_1-\Gamma^5\right)}
\left(\Gamma^2-\Phi_6e^{2\Gamma^5}\right)^2+\frac{1}{2r^2}
e^{-4\left(\Phi_1-\Gamma^5\right)}
\left(\Gamma^3-\Phi_7e^{-4\Gamma^5}\right)^2\right], \nonumber\ee
\be \t T_{16}\left(\Gamma,\Phi\right)=-4r
\left(\Phi_4+\Gamma^8\right)e^{-\left(\Phi_4+\Gamma^8\right)},
\nonumber\ee\be \t T_{17}\left(\Gamma,\Phi\right)=2r^2
e^{-\left(\Phi_4+\Gamma^8\right)}\left(\Phi_1+\Gamma^5\right)^2,
\nonumber\ee\be \t
T_{18}\left(\Gamma,\Phi\right)=\frac{r^2}{2}e^{-\left(\Phi_4+\Gamma^8\right)}
e^{4\left(\Phi_1+\Gamma^5\right)}\left(\Phi_3+\Gamma^7\right)^2.
\nonumber\ee

Each one of the expressions $\t T_{1i}\left(\Gamma,\Phi\right)$
satisfies the equation (\ref{FitT1}). Now as in the previous case,
we make use of the expression for $\t
T_{1}\left(\Gamma,\Phi\right)$ in order to write down a solution
for the equation (\ref{FitT'}). Thus, the solution is given by the
following relations

\be \t
T_1'\left(\Gamma(r,t),\Gamma'(r,t),\Phi(r,t),\Phi'(r,t)\right)=
\sum_{i=1}^8\t
T_{1i}'\left(\Gamma(r,t),\Gamma'(r,t),\Phi(r,t),\Phi'(r,t)\right),\label{tT1i'}
\ee where \be \t T_{11}'\left(\Gamma,\Gamma',\Phi,\Phi'\right)=
\frac{1}{8r^2}\left[
\left(\Gamma^4\right)^2e^{-\left(\Phi_4+\Gamma^8\right)}
+\frac{1}{2}\Phi_8e^{-\left(\Phi_4+\Gamma^8\right)}\Phi_8\right.+
\nonumber\ee \be
\left.\Gamma^4\left(\Phi_8e^{-\left(\Phi_4+\Gamma^8\right)}+
e^{-\left(\Phi_4+\Gamma^8\right)}\Phi_8\right)+\frac{1}{4}
\left(\Phi_8^2e^{-\left(\Phi_4+\Gamma^8\right)}+
e^{-\left(\Phi_4+\Gamma^8\right)}\Phi_8^2\right)\right]-
\nonumber\ee \be
\frac{1}{4r}\left[\Gamma^4{\Gamma^4}'e^{-\left(\Phi_4+\Gamma^8\right)}
+\frac{1}{4}\left(\Phi_8'e^{-\left(\Phi_4+\Gamma^8\right)}\Phi_8+
\Phi_8e^{-\left(\Phi_4+\Gamma^8\right)}\Phi_8'\right)\right.+
\nonumber\ee \be
\frac{{\Gamma^4}'}{2}\left(\Phi_8e^{-\left(\Phi_4+\Gamma^8\right)}+
e^{-\left(\Phi_4+\Gamma^8\right)}\Phi_8\right)+
\frac{\Gamma^4}{2}\left(\Phi_8'e^{-\left(\Phi_4+\Gamma^8\right)}+
e^{-\left(\Phi_4+\Gamma^8\right)}\Phi_8'\right)+ \nonumber\ee \be
\left.\frac{1}{4}
\left(\Phi_8'\Phi_8e^{-\left(\Phi_4+\Gamma^8\right)}+
e^{-\left(\Phi_4+\Gamma^8\right)}\Phi_8'\Phi_8\right)\right],\nonumber\ee
\be \t T_{12}'\left(\Gamma,\Gamma',\Phi,\Phi'\right)=
\frac{1}{4r}\left[
\left(\Gamma^4\right)^2C_1+\Gamma^4\left(\Phi_8C_1+C_1\Phi_8\right)+\frac{1}{2}
\left(\Phi_8C_1\Phi_8+\Phi_8^2C_1+C_1\Phi_8^2\right)\right],
\nonumber\ee where
$C_1=\left(\Phi_4'+{\Gamma^8}'\right)e^{-\left(\Phi_4+\Gamma^8\right)}$;
\be \t
T_{13}'\left(\Gamma,\Gamma',\Phi,\Phi'\right)=-\frac{1}{4r}\left[
\Gamma^4e^{-\left(\Phi_4+\Gamma^8\right)}
+\frac{1}{2}\left(e^{-\left(\Phi_4+\Gamma^8\right)}\Phi_8+
\Phi_8e^{-\left(\Phi_4+\Gamma^8\right)}\right)\right]\times
\nonumber\ee \be
\left[\Gamma^1{\Gamma^5}'+\left(\Gamma^1\Phi_1'-{\Gamma^5}'\Phi_5\right)-
\frac{1}{2}\left(\Phi_1'\Phi_5+\Phi_5\Phi_1'\right)\right],
\nonumber\ee \be \t
T_{14}'\left(\Gamma,\Gamma',\Phi,\Phi'\right)=-\frac{1}{4r}e^{4\Gamma^5}
\left[\Gamma^4e^{-\left(\Phi_4+\Gamma^8\right)}
+\frac{1}{2}\left(e^{-\left(\Phi_4+\Gamma^8\right)}\Phi_8+
\Phi_8e^{-\left(\Phi_4+\Gamma^8\right)}\right)\right]\times
\nonumber\ee \be
\left[\Gamma^3{\Gamma^7}'+\left(\Gamma^3\Phi_3'-e^{-4\Gamma^5}
{\Gamma^7}'\Phi_7\right)-
\frac{1}{2}e^{-4\Gamma^5}\left(\Phi_7\Phi_3'+\Phi_3'\Phi_7\right)\right],
\nonumber\ee \be \t
T_{15}'\left(\Gamma,\Gamma',\Phi,\Phi'\right)=\t
T_{15}\left(\Gamma,\Phi\right), \nonumber\ee \be \t
T_{16}'\left(\Gamma,\Gamma',\Phi,\Phi'\right)=-4r
\left(\Phi_4'+{\Gamma^8}'\right)e^{-\left(\Phi_4+\Gamma^8\right)},
\nonumber\ee \be \t
T_{17}'\left(\Gamma,\Gamma',\Phi,\Phi'\right)=2r^2
e^{-\left(\Phi_4+\Gamma^8\right)}\left(\Phi_1'+{\Gamma^5}'\right)^2,
\nonumber\ee \be \t
T_{18}'\left(\Gamma,\Gamma',\Phi,\Phi'\right)=\frac{r^2}{2}e^{-\left(\Phi_4+\Gamma^8\right)}
e^{4\left(\Phi_1+\Gamma^5\right)}\left(\Phi_3'+{\Gamma^7}'\right)^2.
\nonumber\ee

Now we are in position to calculate the fermion generating
operator $\Omega$. In order to achieve this aim, let us recall
that we have the following set of irreducible first class
constraints \be T\equiv\left(\ba{c} \t T_a'\cr\T_A\ea\right),
\qquad a=1,2;\qquad A=1,2,...,8; \label{1cc} \ee whose expressions
are given in the equations (\ref{Texplicit}), (\ref{tT2'simp}) and
(\ref{tT1i}). We put into correspondence with the complete set of
irreducible first class constraints (\ref{1cc}) the following
ordered operator pairs \be \left(\ba{c} \t
T_a'\cr\T_A\ea\right)\longrightarrow \left(\ba{cc}
{\lambda^a}'(r,t),&{\pi^a}'(r,t)\cr
{\lambda^A}(r,t),&{\pi_A}(r,t)\ea\right), \left(\ba{cc}
{C^a}'(r,t),&{\overline \P_a}'(r,t)\cr {C^A}(r,t),&{\overline
\P_A}(r,t)\ea\right), \left(\ba{cc} {\P^a}'(r,t),&{\overline
C_a}'(r,t)\cr {\P^A}(r,t),&{\overline C_A}(r,t)\ea\right),
\label{ghost}\ee  where the active Lagrange multipliers read
${\lambda^1}'=N^\perp$, ${\lambda^2}'=A_0$. The ghost numbers and
the statistics of these magnitudes read \be gh(C)=gh(\P)=1, \qquad
gh(\overline C)=gh(\overline \P)=-1,\nonumber \ee \be
\eps(\lambda)=\eps(\pi)=0, \qquad \eps(C)=\eps(\overline
C)=\eps(\P)=\eps(\overline \P)=1. \nonumber\ee

The following supercommutators are nonzero \be
[\lambda^j(r,t),\pi_j(r',t)]=i\b h\delta(r-r'), \nonumber\ee \be
[C^j(r,t),\overline\P_j(r',t)]=i\b h\delta(r-r'), \label{scomm}\ee
\be [\P ^j(r,t),\overline C_j(r',t)]=i\b h\delta(r-r'),
\nonumber\ee while all remaining supercommutators of these
quantities vanish, i.e.,
$[C^i(r,t),C^j(r',t)]=0$,
$[\overline\P^i(r,t),\overline\P_j(r',t)]=0$, etc., where
$i,j=a,A$.

We have assumed as well that the following relations take place
\be C^i(r,t)C^i(r',t)\delta(r-r')=\overline C_i(r,t)\overline
C_i(r',t)\delta(r-r')=0, \nonumber\ee \be
\P^i(r,t)\P^i(r',t)\delta(r-r')=
\overline\P_i(r,t)\overline\P_i(r',t)\delta(r-r')=0.
\label{ccdelta}\ee

Thus, the fermion generating operator $\Omega$ adopts the form
given by the expression (\ref{Omega}) \be
\Omega=\Omega'(\Gamma,\Gamma',\Phi,\Phi',C',\overline{\P}')\!+\!
\T_A(\Gamma,\Pi,\Phi)C^A\!+\!\Pi'_a\P'^a\!+\!\pi_A\P^A,\quad
a=1,2;\ \ A=1,2,...8; \label{Omegaemda}\ee where the the Fermi
operator $\Omega'$ obeys the following relations
$[\Omega,\Omega']=[\Omega',\T_A]=0$, possesses the ghost number
$\,\,gh(\Omega')=1$, and satisfies the boundary condition
$\Omega'(\Gamma,\Gamma',\Phi,\Phi',C',0)=\t
T_a'(\Gamma,\Gamma',\Phi,\Phi')C'^a$.

We try to find the solution for the equation (\ref{Omegaemda}) in
the following form (here we use a Weyl basis) \be \Omega'\!=\!\t
T_a'(\Gamma,\Phi,\Phi'){C'}^a\!+\!\t
U_{bc}^a(\Gamma,\Phi,\Phi')\!\left[\overline\P'_aC'^c\left(\frac{\pa}{\pa
r}C'^b\right)\!+\!\left(\frac{\pa}{\pa
r}C'^b\right)\overline\P'_aC'^c\!+\!C'^c\left(\frac{\pa}{\pa
r}C'^b\right)\overline\P'_a\right]\!\!. \label{Omega'}\ee

By keeping only terms with lowest number of ghost ($2$--ghost) we
obtain \be [\Omega,\Omega']=\left[\t T'_a(r,t),\t
T'_b(r',t)\right]C'^a(r,t)C'^b(r',t)+ \nonumber\ee \be 6i\b h
U_{bc}^a(r,t)\t T'_a(r,t)C'^c(r,t)\left(\frac{\pa}{\pa
r}C'^b(r,t)\right)\delta(r-r')+\label{OmOm'}\ee \be 9\b
h^2\delta(r-r')\left(\frac{\pa}{\pa r'}\delta(r'-r)\right)
\left[U_{bj}^i(r,t),U_{if}^j(r',t)\right]\left(\frac{\pa}{\pa
r}C'^b(r,t)\right)C'^f(r',t)+\nonumber\ee \be 4\!-\!\mbox{\rm
ghost terms}+ 6\!-\!\mbox{\rm ghost terms}. \nonumber\ee

In our case the following commutation relations for the
constraints $\t T'_a(r,t)$ take place $\left[\t T'_1(r,t),\t
T'_2(r',t)\right]=\left[\t T'_2(r,t),\t T'_2(r',t)\right]=0$,
while the commutator $\left[\t T'_1(r,t),\t T'_1(r',t)\right]$ can
be written in the form \be \left[\t T'_1(r,t),\t
T'_1(r',t)\right]=i\b h F_1(r)\frac{\pa}{\pa r'}\delta(r'-r)+i\b h
F_2(r)\delta(r-r').\label{comF1F2}\ee

Thus, we have \be \left[\t T'_a(r,t),\t T'_b(r',t)\right]C'^a(r,t)
C'^b(r',t)=i\b h F_1(r)C'^1(r,t)C'^1(r',t)\frac{\pa}{\pa
r'}\delta(r'-r)=\nonumber\ee \be -i\b h
F_1(r)C'^1(r,t)\left(\frac{\pa}{\pa
r}C'^1(r,t)\right)\delta(r-r').\label{tTa'tTb'}\ee

The exact form of the function $F_1(r)$ and possible solutions of
(\ref{Omegaemda}) are currently under investigation.

\section{Conclusions}

The canonical quantization of dynamical systems introduced by I.A.
Batalin and E.S. Fradkin in \cite{bf} and I.A. Batalin, E.S.
Fradkin  and T.E. Fradkina in \cite{bff} is applied to the
four--dimensional Einstein--Maxwell Dilaton--Axion theory.

By performing an ADM decomposition of the metric and considering
the spherically symmetric anzats with radial fields, the total
Lagrangian density of the theory (gravity coupled to matter
fields) is written as an expression first order in time
derivatives of the fields which defines a curved phase space with
two irreducible first class constraints. Thus, the canonical
quantization method mentioned above can be applied to this
effective system. However, in order to achieve thin goal, some
generalizations of the method have been performed.

This paper actually constitutes the first part of our
investigation because the explicit form of the fermion generating
operator $\Omega$ and the total unitarizing Hamiltonian of the
theory $H$ are not included. This is the subject of our current
research activity.

\section*{Acknowledgments}

One of the authors (AHA) is really grateful to the Theoretical
Physics Department of the Aristotle University of Thessaloniki for
providing a stimulating atmosphere while this investigation was
carried out. He also acknowledges a grant for postdoctoral studies
provided by the Greek Government. The research of AHA was
supported by grants CIC-UMSNH-4.18 and CONACYT-42064-F.

\newpage

\section*{Appendix A}

For any continuous differential function $f(z)$ the following
identity holds \be
f(z)\delta'(z-z_0)=f(z_0)\delta'(z-z_0)-f'(z_0)\delta(z-z_0),
\label{A1}\ee where now the prime denotes derivatives with respect
to the variable $z$.

The space derivative of the commutator of two $\Phi$ fields is
given by the following relation

\be [\Phi_a(r,t),\Phi_b'(r',t)]=-i\b
h\,\omega_{ab}^{(0)}\frac{\pa}{\pa r'}\delta(r'-r).\label{A2}\ee

One of the identities of Ref. \cite{suzuki} is useful for the case
we are considering, namely, the following one \be
e^{\lambda\left(A+B\right)}= e^{\left(\lambda/2\right)A}e^{\lambda
B}e^{\left(\lambda/2\right)A}+ \frac{1}{2}\int_0^\lambda
dt\int_0^t ds\ e^{\left(t/2\right)A}
e^{tB}G(s)\,e^{\left(t/2\right)A}e^{\left(\lambda-t\right)\left(A+B\right)},
\label{A3}\ee where \be G(s)=\int_0^s
du\left\{\frac{1}{2}e^{\left(u/2\right)A}\left[A,\left[A,B\right]\right]
e^{-\left(u/2\right)A}+e^{-uB}\left[B,\left[A,B\right]\right]
e^{uB}\right\}. \nonumber\ee In our case the function $G(s)=0$. We
also set $\lambda=1$. Finally, we should point out that this
formula is used just in the case of noncommuting $\Phi_a$ fields.

\newpage


\end{document}